# On Stability of Non-inflectional Elastica

Milan Batista

University of Ljubljana, Faculty of Maritime Studies and Transport

Pot pomorščakov 4, 6320 Portorož, Slovenia

milan.batista@fpp.uni-lj.si

**Abstract**

This study considers the stability of a non-inflectional elastica under a conservative end force subject to the Dirichlet, mixed, and Neumann boundary conditions. It is demonstrated that the non-inflectional elastica subject to the Dirichlet boundary conditions is unconditionally stable, while for the other two boundary conditions, sufficient criteria for stability depend on the signs of the second derivatives of the tangent angle at the endpoints.

## 1. Introduction

In recent years, Lessinnes and Goriely [1], among others, have considered the stability of equilibria for initially uniformly curved weightless inextensible and unshearable rods (hereafter referred to as elastica), subject to Dirichlet and Neumann boundary conditions. For each boundary condition, they defined an index, the value of which determines the stability or instability of equilibria. This value can be determined by a geometric analysis of the phase-plane trajectories of stationary solutions.

This study demonstrates that, for non-inflectional elastica [2]; that is, elastica bent in one direction, the stability of equilibria may in certain cases be determined only from the signs of the second derivative of the equilibria at endpoints. Using results

19.12.2018 9:16

from [1] we can also extend sign criteria to prove the instability of a non-inflectional elastica subject to the Neumann boundary condition.

## 2. Equilibrium

A planar elastica with one end spatially fixed and another spatially free will be considered, subject to a conservative force $F$. The potential energy $\Pi$ of such elastica is commonly expressed in the following dimensionless form:

$$\Pi[\theta] = \int_0^1 \left[ \tfrac{1}{2}(\theta' - a)^2 + \omega^2 \cos(\theta + \alpha) \right] ds \tag{1}$$

where $s \in [0,1]$ is the normalized arc length parameter, $(\ )' = d(\ )/ds$, $\theta = \theta(s)$ is the angle between the $x$-axis and the tangent to the curve, $\alpha$ is the angle between the force and the negative direction of the $x$-axis, $a > 0$ is the reference dimensionless curvature, and $\omega \geq 0$ is the load parameter. Because only non-inflectional elastica was considered; therefore, $\theta$ is subject to the condition

$$\theta'(s) > 0, \ s \in [0,1] \tag{2}$$

To determine the equilibrium shape (or configuration or state) of the elastica, we make $\delta\Pi[\theta] = 0$. In this manner, we deduce the equilibrium equation and boundary conditions for $\theta$:

$$\theta'' + \omega^2 \sin(\theta + \alpha) = 0 \tag{3}$$

$$(\theta' - a)\delta\theta \Big|_{s=0} = 0, \quad (\theta' - a)\delta\theta \Big|_{s=1} = 0 \tag{4}$$

where $\delta\theta$ is variation of $\theta$. The solution of (3); that is, the equilibria of $\Pi$, subject to (2) is expressed as [2-4]

$$\theta = -\alpha + 2\operatorname{am}(k\,\omega s + C, k^{-1}) \tag{5}$$

where *am* is Jacobi's amplitude function [5], $k \geq 1$ is the elliptic modulus, and $C$ is an integration constant. Because *am* is periodic with period $2K$, $C$ can always be selected to lie within the interval

$$-K \leq C < K, \tag{6}$$

where $K = K(k^{-1})$ is the elliptic integral of the first type. The first and second derivatives of (5) are

$$\theta' = 2k\,\omega\,\mathrm{dn}\left(k\,\omega s + C, k^{-1}\right) \tag{7}$$

$$\theta'' = -2\omega^2\,\mathrm{sn}\left(k\,\omega s + C, k^{-1}\right)\mathrm{cn}\left(k\,\omega s + C, k^{-1}\right), \tag{8}$$

where *sn*, *cn*, and *dn* are Jacobi's elliptic functions [5].

The boundary conditions (4) for $\theta$ are now considered. Each can be satisfied in two manners, as follows.

- If $\delta\theta = 0$, and at the endpoint $\theta$ is prescribed, the endpoint can be said to be clamped.
- If $\theta' = a$, the endpoint can be said to be pinned or hinged. In this case, $\delta\theta' = 0$.

From the above, the following types of boundary conditions can be established.

- Dirichlet:

$$\theta(0) = \theta_0, \quad \theta(1) = \theta_1 \tag{9}$$

- Mixed:

  (a)
  $$\theta(0) = \theta_0, \quad \theta'(1) = a \tag{10}$$

  (b)
  $$\theta'(0) = a, \quad \theta(1) = \theta_1 \tag{11}$$

- Neumann:

$$\theta'(0) = a, \quad \theta'(1) = a \tag{12}$$

This section can be concluded with the following lemma.

19.12.2018 9:16

**Lemma 1.** *If a non-inflectional elastica is subject to Neumann boundary conditions, the signs of the second derivatives of θ at the endpoints are either equal or opposite.*

*Proof.* From (7) and (12), the equation obtained is

$$\mathrm{dn}\left(C, k^{-1}\right) = \mathrm{dn}\left(k\,\omega + C, k^{-1}\right) \tag{13}$$

Because the function $dn$ is periodic with period $2K$ and symmetric, this equation is identically satisfied in two cases:

(a) 
$$k\,\omega = 2nK \tag{14}$$

(b) 
$$k\,\omega = -2C + 2nK, \tag{15}$$

where $n \geq 0$ is an integer. By substituting these solutions into expression (8) for the second derivative of $\theta$, and taking into account the symmetry and periodicity of the functions $sn$ and $cn$, we obtain

$$\theta''(1) = \theta''(0) \quad \text{(solution (a))} \tag{16}$$

$$\theta''(1) = -\theta''(0) \quad \text{(solution (b))} \tag{17}$$

This lemma implies that the situation in which $\theta''$ vanishes only at one endpoint is impossible for Neumann boundary conditions; it can be zero only at both ends simultaneously. It is noted that this is the case when the non-inflectional elastica forms closed loops.

### 3. Stability

To study the stability of solution (5), the second variation $\delta^2 \Pi[\theta]$ must be examined. It is well known [6] that the equilibrium is stable if $\delta^2 \Pi[\theta] > 0$ for any admissible variation $\delta\theta$; that is, a nontrivial variation satisfying given boundary conditions (see below) [7].

From (1), it can be seen that

$$\delta^2 \Pi[\theta] = \int_0^1 \left( \delta\theta'^2 - \omega^2 \cos(\theta + \alpha) \delta\theta^2 \right) ds \tag{18}$$

By differentiating (3) with respect to $s$, we obtain $\theta''' + \omega^2 \cos(\theta + \alpha)\theta' = 0$. Owing to (2), this can always be solved for $\cos(\theta + \alpha)$ and then the resulting expression can be substituted into (18). Upon integration by parts, we obtain

$$\delta^2 \Pi[\theta] = \frac{\theta''}{\theta'} \delta\theta^2 \bigg|_0^1 + \int_0^1 \left( \delta\theta' - \frac{\theta''}{\theta'} \delta\theta \right)^2 dx \ . \tag{19}$$

For future use, this expression is rewritten as

$$\delta^2 \Pi[\theta] = B + J \tag{20}$$

where

$$B \equiv \frac{\theta''}{\theta'} \delta\theta^2 \bigg|_1 - \frac{\theta''}{\theta'} \delta\theta^2 \bigg|_0 \quad \text{and} \quad J \equiv \int_0^1 \left( \delta\theta' - \frac{\theta''}{\theta'} \delta\theta \right)^2 dx \tag{21}$$

It is clear that $J \geq 0$ for any $\delta\theta$. Therefore, $\delta^2 \Pi > 0$ is obtained when either of the following conditions are fulfilled:

a) $J = 0$ and $B > 0$ or

b) $J > 0$ and $B > -J$.

Hereafter, the conditions imposed on $B$ must be fulfilled for any admissible $\delta\theta$.

The question of the stability of equilibria is thus reduced to an investigation of the signs of $B$ and $J$, subject to the following boundary conditions.

- Dirichlet boundary conditions:

$$\delta\theta(0) = \delta\theta(1) = 0 \tag{22}$$

- Mixed boundary conditions:

(a) $$\delta\theta(0) = 0, \quad \delta\theta'(1) = 0 \tag{23}$$

(b) $$\delta\theta'(0) = 0, \quad \delta\theta(1) = 0 \tag{24}$$

- Neumann boundary conditions:

$$\delta\theta'(0) = \delta\theta'(1) = 0 \tag{25}$$

We firstly consider $J$. From (21), it can be seen that $J = 0$ only when $\delta\theta' - \dfrac{\theta''}{\theta'}\delta\theta = 0$; that is, when

$$\delta\theta = c\,\theta', \tag{26}$$

where $c$ is a constant [8, 9]. The boundary conditions (22) to (25) for which the variation (26) is admissible are now considered.

**Lemma 1.** *If the non-inflectional elastica has at least one endpoint clamped, $\delta\theta = c\,\theta'$ is not an admissible variation.*

*Proof.* Suppose that $\delta\theta = c\,\theta'$ is an admissible variation. By assumption, $\theta' \neq 0$; therefore, the boundary condition $\delta\theta = 0$ can be satisfied only if $c = 0$. However, in this case, (26) becomes $\delta\theta \equiv 0$ (no variation), which is not an admissible variation. This contradiction proves the lemma.

**Lemma 2.** *If the non-inflectional elastica has at least one endpoint pinned, and if at that point $\theta'' \neq 0$, $\delta\theta = c\,\theta'$ is not an admissible variation.*

*Proof.* Let $\delta\theta = c\,\theta'$ be an admissible variation; then, $\delta\theta' = c\,\theta''$. At the pinned endpoint, it is required that $\delta\theta' = 0$. But this can be satisfied only if $c = 0$, because at that point $\theta'' \neq 0$ by assumption. Therefore, $\delta\theta \equiv 0$, which is not an admissible variation.

**Lemma 3.** *If the non-inflectional elastica is subject to Neumann boundary conditions and if $\theta''(0) = \theta''(1) = 0$, there exists an admissible variation so that $\delta^2\Pi[\theta] = 0$.*

*Proof.* Let $\delta\theta = c\,\theta'$. Then, $\delta\theta' = c\,\theta''$ and this is zero at the endpoints for any $c \neq 0$ because of the assumption $\theta''(0) = \theta''(1) = 0$. Therefore, $\delta\theta = c\,\theta'$ is an admissible variation for which $B = 0$ and $J = 0$. Hence, $\delta^2\Pi[\theta] = 0$.

Thus, it has been inferred that $J > 0$ for all of the boundary conditions, except for the Neumann boundary conditions with $\theta'' = 0$ at both endpoints, in which case $J = 0$ and $B = 0$. The latter means that condition (a) cannot occure. Thus, only case (b) remains in the investigation; that is, with the condition $B > -J$. Now, in expression (21) for $B$, the only signed variable is $\theta''$, because, by assumption, $\theta' > 0$. However, the sign of $B$ also depends on the magnitudes of $\dfrac{\theta''}{\theta'}\delta\theta^2$ at each end, particularly the magnitude of $\delta\theta^2$. However, the condition $B > -J$ will certainly be fulfilled if $B \geq 0$. The requirements $J > 0$ and $B \geq 0$ are therefore sufficient for the stability of equilibria. The above discussion is summarized with the following corollary.

**Corollary 1.** *If $B \geq 0$, the equilibrium is stable for all of the boundary conditions except for the Neumann boundary conditions, with $\theta''(0) = \theta''(1) = 0$.*

The sufficient conditions for the shape stability of the non-inflectional elastica are thus governed only by the sign of the boundary term $B$. These must be established separately for each type of boundary condition. Based on Corollary 1, we can state the following theorems, which can be trivially proven by verification.

**Theorem 1.** *The equilibrium shapes of a non-inflectional elastica subject to the Dirichlet boundary conditions are stable.*

In this theorem, it is noted that there is no stability condition, so this is the ultimate result concerning the Dirichlet boundary condition. It was provided for the initially straight rod by Born [10]. The same conclusion can also be inferred from the Lessinnes and Goriely geometric criteria [1].

**Theorem 2a.** *The equilibrium shape of a non-inflectional elastica subject to mixed end conditions* (10) *is stable if*

$$\theta''(1) \geq 0 \quad . \tag{27a}$$

**Theorem 2b.** *The equilibrium shape of the non-inflectional elastica subject to mixed end conditions* (11) *is stable if*

$$\theta''(0) \leq 0 \quad (27)\text{b}$$

**Theorem 3.** *The equilibrium shape of a non-inflectional elastica subject to Neumann boundary conditions is stable if*

$$\theta''(0) < 0 \quad \text{and} \quad \theta''(1) > 0 \quad (28)$$

Several examples of stable non-inflectional elastica shapes, together with phase portraits and bifurcation diagrams, are illustrated in Figs. 1 to 9. In these figures,

$$q \equiv a/2\pi \quad \text{and} \quad \beta \equiv \left(\omega^2/\pi\right) \quad (29)$$

where $q$ is the loops number and $\beta$ is the multiplier of the Euler critical force that buckles a hinged straight rod. These theorems allow for easy identification of stability regions in the bifurcation diagrams (Figs. 3, 6, and 8); stability regions are determine by (27)-(28) using (8). Fig 2,4,5,7,9 shows elastica shape correspond to points on the bifurcation diagrams and also a phase plane diagram where one can decide stability of the shape according to Lessinnes-Goriely criterion [1] (see below).

## 4. Instability

This section proves a theorem concerning the instability of a non-inflectional elastica subject to the Neumann boundary condition. For this purpose, we use the Lessinnes-Goriely criterion [1] (theorem 3 therein): if $V = -\omega^2 \cos(\theta + \alpha)$, the equilibrium is stable if it crosses more maximum than minimum of $V$, and it is unstable if it crosses more minimum than maximum. The case of an equal number of maximum and minimum is not covered. In the case of the non-inflectional elastica $\theta' = 2\omega\sqrt{k^2 - \sin^2(\theta + \alpha)}$, the maximum and minimum of $V$ correspond to the minimum of $\theta'$, and respectively the maximum of $\theta'$. This observation allows us to apply the criteria in the following form:

*If, along a non-inflectional elastica, the number of minima of $\theta'$ is less than the number of its maxima, the equilibrium is unstable.*

19.12.2018 9:16

**Theorem 4.** *If $\theta''(0) > 0$ and $\theta''(1) < 0$, the equilibrium shapes of a non-inflectional elastica subject to the Neumann boundary conditions are unstable.*

*Proof.* This case is covered by solution (15). If the variable $s$ is changed to $\sigma \equiv k\omega s$, $\vartheta(\sigma) \equiv \theta(s(\sigma))$ is defined for $C \leq \sigma \leq -C + 2nK$. Thus, $\vartheta'(\sigma)$ and $\vartheta''(\sigma)$ become periodic with the period $2K$ (Fig. 10). Next, it is observed that the assumption $\vartheta''(C) > 0$ is fulfilled if $-K < C < 0$. In this case, we also have $\vartheta''(-C + 2nK) = \vartheta''(-C) < 0$, as required by the theorem. The extremes of $\theta'$ are now examined for three intervals. The first interval is $-K < C \leq \sigma < 0$. Here, $\vartheta'$ has no extrema. The next interval $0 \leq \sigma < 2nK$ consists of $n$ periods. In each period, there is one maximum and one minimum of $\vartheta'$. The last interval $2nK \leq \sigma \leq -C + 2nK < K + 2nK$ has one maximum of $\vartheta'$. The number of maxima is thus one greater than the number of minima. Hence, the elastica is unstable.

It is noted that, by similar reasoning, Theorem 3 can be proved without appealing to Corollary 1.

## 5. Conclusion

This study demonstrated that the shape of a non-inflectional elastica subject to Dirichlet boundary conditions is stable. For a non-inflectional elastica subject to mixed conditions, two theorems that provide sufficient conditions for its stability were established. For a non-inflectional elastica subject to Neumann boundary conditions that does not form closed periods, a sufficient condition for its stability as well as its instability was provided.

When investigating the stability of a non-inflectional elastica, the method proposed by this study is slightly simpler than that suggested in [1] because no trajectory has to be examined in the phase space; only the second derivative values at the endpoints must be calculated. This also allows for simple identification of stability regions in bifurcation diagrams associated with particular boundary conditions. However, the proposed method exhibits limitations compared to that provided in [1]:

19.12.2018 9:16

it allows for establishing only sufficient conditions for stability, and cannot be used to establish conditions for instability.

19.12.2018 9:16

**Figures**

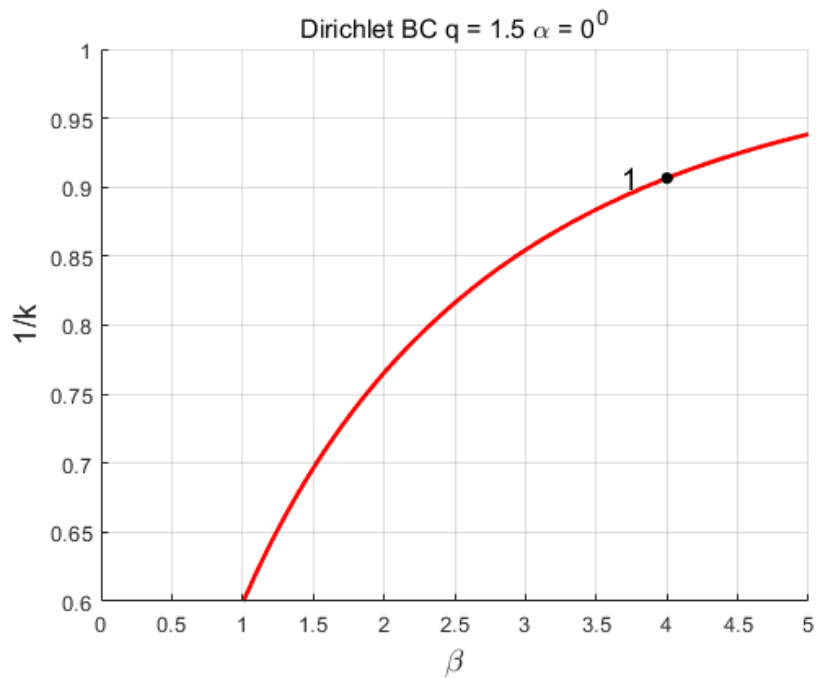

**Figure 1**. Bifurcation diagram of Dirichlet problem.

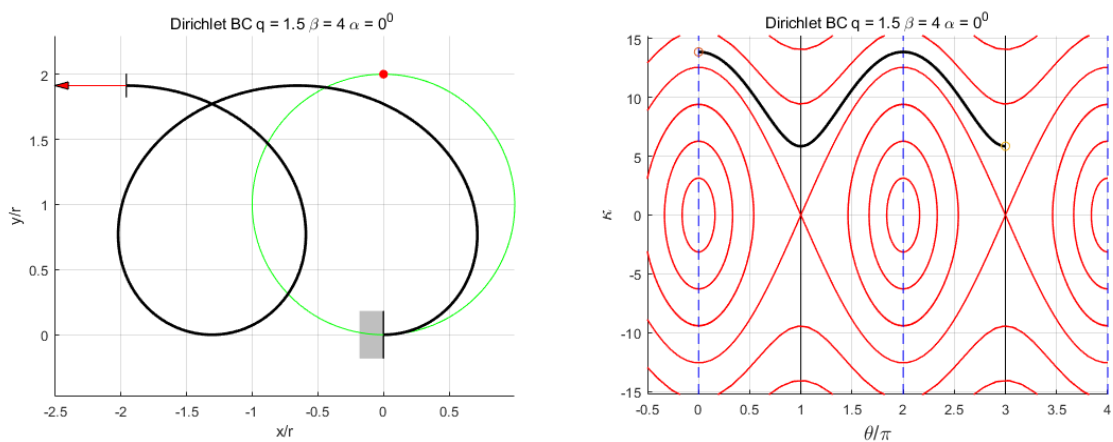

**Figure 2**. Stable forms of non-inflectional elastica corresponding to points 1 in Fig. 1. On bifurcation diagrams the trajectories has equal number of minima and maxima.

19.12.2018 9:16

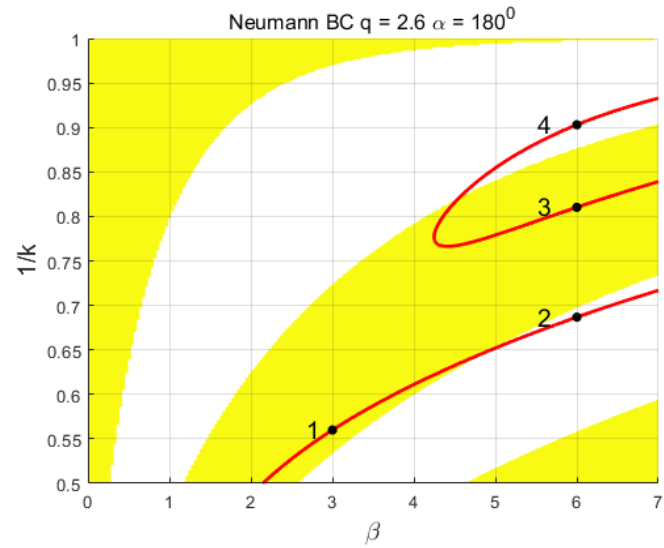

**Figure 3.** Bifurcation diagram of Neumann problem. Stable regions given by (28) are non-shaded; unstable regions are shaded.

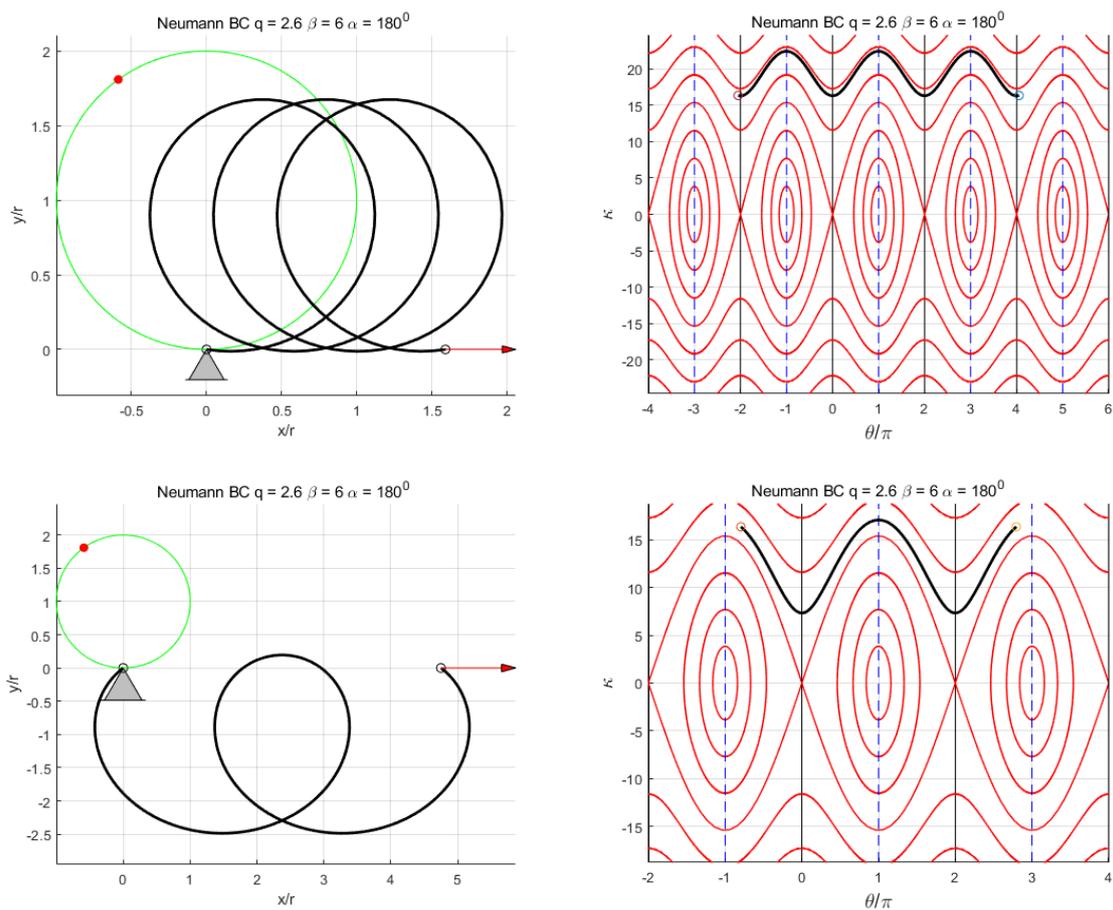

**Figure 4**. Stable forms of non-inflectional elastica corresponding to points 2 and 4 in Fig. 3. On bifurcation diagrams the trajectories has more minima than maxima.

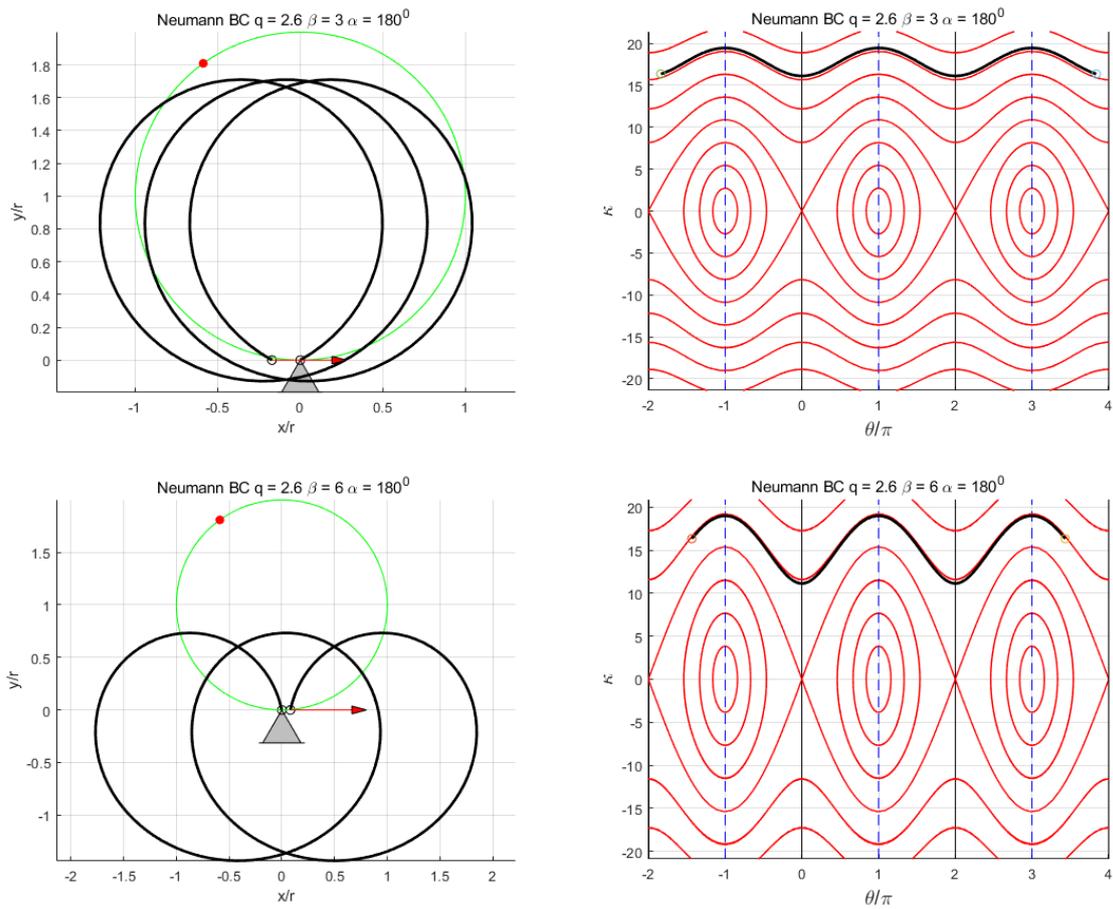

**Figure 5**. Unstable forms of non-inflectional elastica corresponding to points 1 and 3 in Fig. 3. On bifurcation diagrams the trajectories has more maxima than minima.

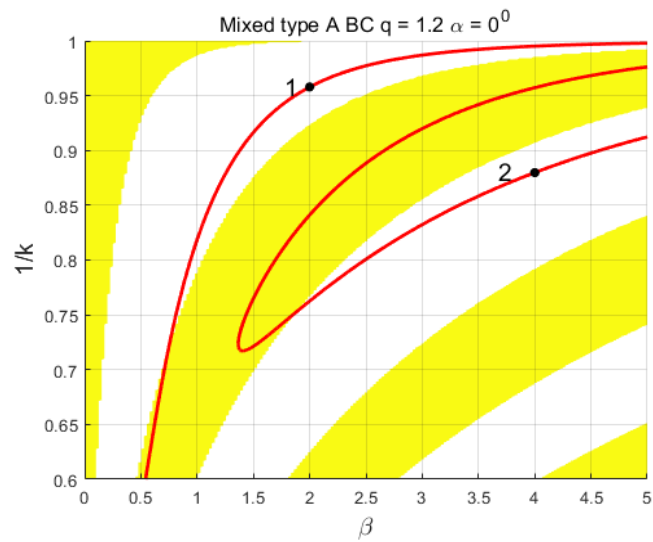

**Figure 6.** Bifurcation diagram of mixed problem. Stable regions given by (27)a are non-shaded;

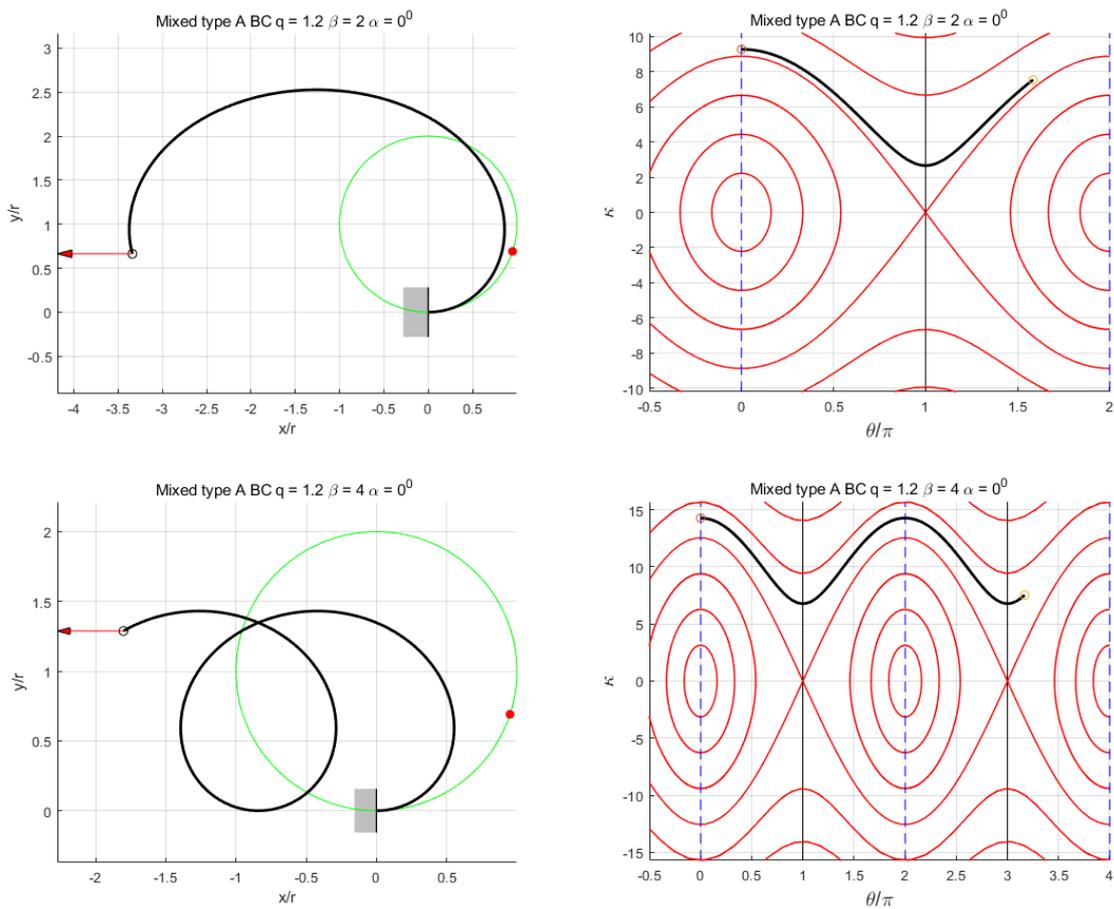

**Figure 7**. Stable forms of non-inflectional elastica corresponding to points in Fig. 6. On bifurcation diagrams the trajectories has equal number of minima and maxima.

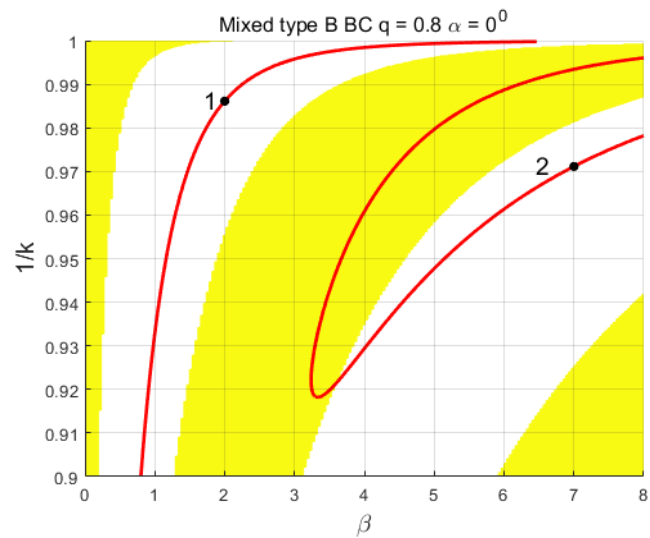

**Figure 8**. Bifurcation diagram of mixed problem. Stable regions given by (27)b are non-shaded;

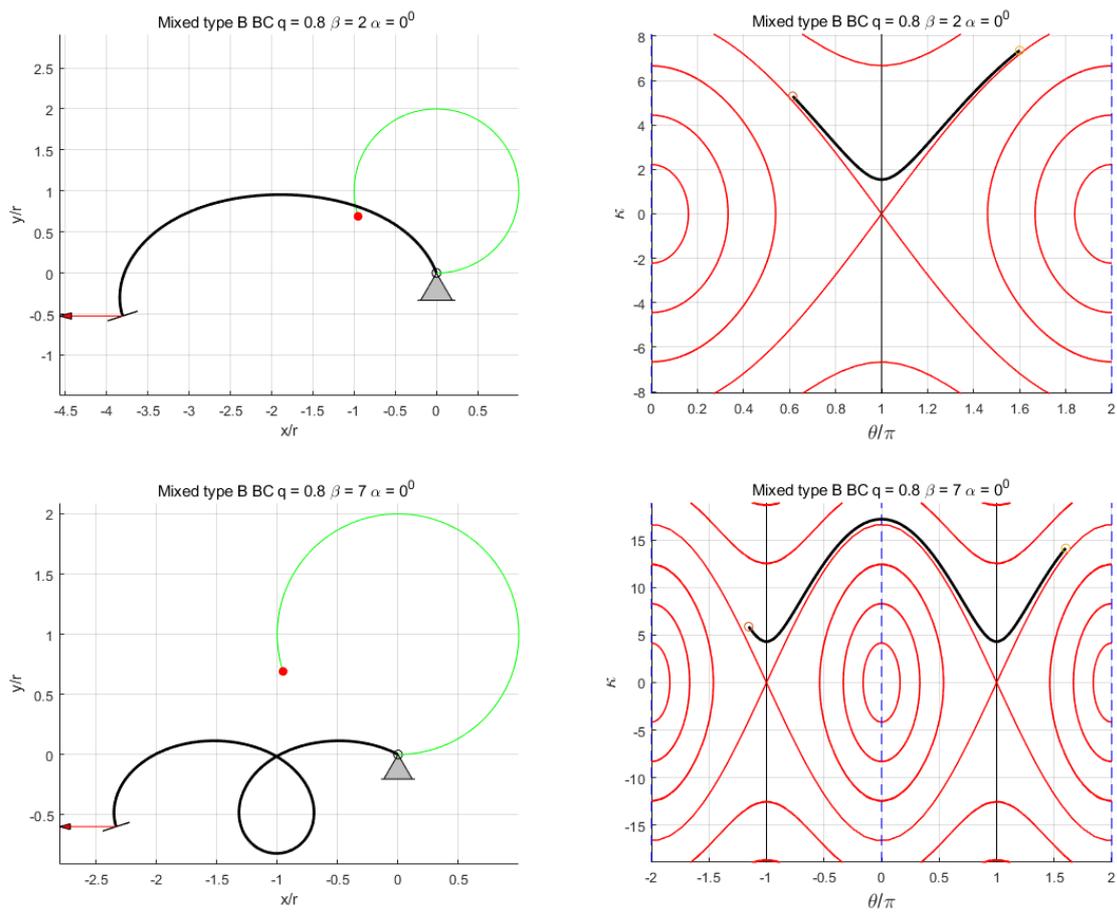

**Figure 9**. Stable forms of non-inflectional elastica corresponding to points in Fig. 8. On bifurcation diagrams the trajectories has more minima than maxima.

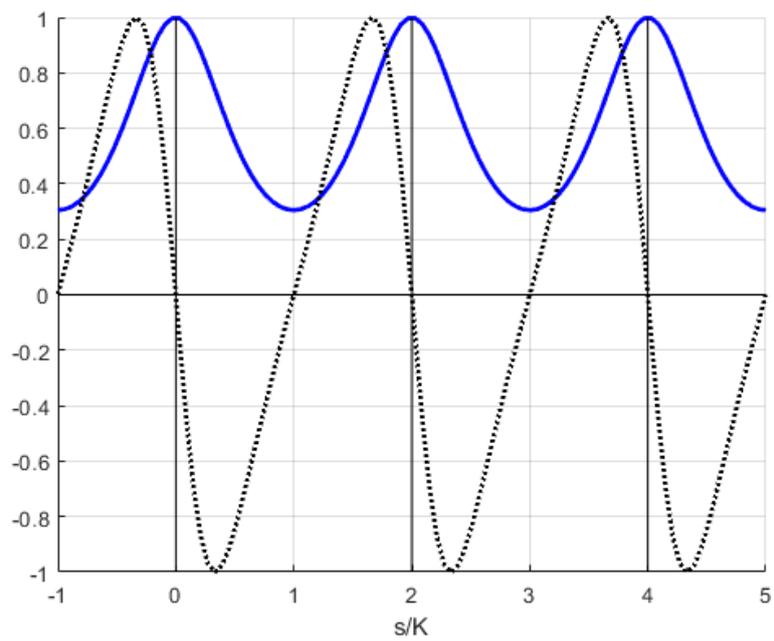

**Figure 10.** Graph of function $dn$ (solid line) and its normalized derivative (dotted line).